\documentclass[traditabstract]{aa}
\usepackage{graphicx,color} 
\usepackage{rotating,subfigure,amssymb,afterpage} 
\usepackage{txfonts}
\usepackage{aalongtable}

\newcommand{\propsim}{\lower 3pt \hbox{$\, \buildrel {\textstyle 
      \propto}\over {\textstyle \sim}\,$}} 
\newfont{\gwpfont}{cmssq8 scaled 1000}
\newcommand{\reflex}{{\gwpfont REFLEX}}
 
\begin{document} 
\title{The ROSAT-ESO Flux Limited 
X-ray Galaxy Cluster Survey (REFLEX II) \\
I. Newly identified X-ray luminous clusters at $z \ge $ 0.2 
\thanks{Based on the data obtained at the European  
Southern Observatory, La Silla, Chile}}
\author{Gayoung Chon\inst{1} and Hans B\"ohringer\inst{1}}
\offprints{Gayoung Chon, gchon@mpe.mpg.de} 
\institute{$^1$ Max-Planck-Institut f\"ur extraterrestrische Physik, 
                 D 85748 Garching, Germany} 
\date{Submitted 01 Sep 2011;} 
  
\abstract 
{We report 19 intermediate redshift clusters newly detected in the ROSAT
All-Sky survey that are spectroscopically confirmed. They form a part of 
911 objects in the \reflex\ II cluster catalogue with a limiting flux of 
1.8$\times10^{-12}$ erg/s/cm$^2$ in the 0.1-2.4 keV ROSAT band at redshift 
$z \ge 0.2$. 
In addition we report three clusters from the \reflex\ III supplementary
catalogue, which contains objects below the \reflex\ II flux limit but
satisfies the redshift constraint above.
These clusters are spectroscopically followed-up by our ESO NTT-EFOSC2 
campaigns for the redshift measurement. We describe our observing and data
reduction methods. 
We show how X-ray properties such as spectral hardness ratio 
and source extent can be used as important diagnostics in 
selecting galaxy cluster candidates.
Physical properties of the clusters are subsequently calculated 
from the X-ray observations. This sample contains the high mass 
and intermediate-redshift galaxy clusters for astrophysical 
and cosmological applications.
}
\authorrunning{Chon and B\"ohringer} 
\titlerunning{X-ray luminous Clusters in REFLEX II}
\maketitle 

\section{Introduction}

Galaxy clusters are the largest well-defined objects
in our Universe characterized by a dynamical equilibrium
configuration (e.g. Navarro, Frenk and White 1995). 
They form an integral part of the cosmic large-scale structure 
and are ideal probes for the testing of cosmological models 
(e.g. Allen et al. 2011).
However, they are also astrophysical superlatives in many
respects causing the largest deflections as gravitational
lenses (e.g. Schneider et al. 1992), major mergers
of galaxy clusters during the large-scale formation constituting
the most energetic events in the Universe apart from 
the big bang (Sarazin 2002), and imprinting a distinct 
frequency-dependent feature
onto the cosmic microwave background by means of the
Sunyaev-Zel'dovich effect (Sunyaev \& Zel'dovich 1970). Therefore,
they are extremely interesting astrophysical laboratories 
to study various important processes on very large scales.

With technological advances in the past few decades, 
the detection of an increasing number of galaxy clusters 
has enabled us to study their properties in detail, and to 
apply the statistics of the cluster population to a
number of cosmological studies.

One of the most efficient methods for detecting galaxy 
clusters as truly gravitationally bound objects in a close 
to mass-selected way is the use of X-ray surveys (B\"ohringer 2008).
In these surveys galaxy clusters are observed as extended
X-ray sources owing to the thermal X-ray emission of the 
very hot intracluster medium (ICM).

The largest sample of X-ray luminous galaxy clusters to date is
the one that has been identified in the ROSAT all-sky
Survey (RASS), the only all-sky X-ray survey yet conducted
with an imaging X-ray telescope (Tr\"umper 1993).
The southern \reflex\ I (B\"ohringer et al. 2004) and 
northern NORAS I (B\"ohringer et al. 2000) cluster
samples identified in the RASS representing a total of 996
X-ray detected galaxy clusters provide the largest
X-ray catalogue compiled, where the \reflex\ survey
has the most well-understood three-dimensional 
survey selection function. 

With the completion of two recent spectroscopic observing campaigns, 
we have almost completed the \reflex\ II survey by 
adding 50-60 newly determined redshifts.
The \reflex\ II survey effectively doubles the number
of galaxy clusters from 447 to 911 objects.

The most massive galaxy clusters provide 
the tightest constraining power for cosmological model
testing and are also the most interesting astrophysical 
laboratories. We therefore communicate here the detailed
properties of 22 of the most massive and highest redshift 
clusters ($z \ge 0.2$) among the sample of 50-60 X-ray 
detected clusters with newly determined redshifts 
from our recent observing campaigns. Our major aim 
in this paper is to communicate the data of newly 
discovered X-ray luminous clusters and most importantly 
to provide the redshift information for all the cluster 
galaxies with new redshift measurements, so that they can 
enter public data bases (e.g. NED, SIMBAD) and be used by 
the community.

In this paper, we provide a brief description of \reflex\ II
and the present cluster sample in Section 2. 
Our optical observations and data reduction is outlined 
in Section 3 and in Section 4 the X-ray properties and their 
use as the diagnostics of the nature of the sources are described.
In the latter section we also provide some comments on individual 
clusters. Section 5 then provides a summary and some conclusions
about the application of these findings.

\section[]{Description of the cluster sample}

The galaxy clusters of the present sample are the last 
galaxy cluster candidates of the \reflex\ II
catalogue for which a final confirmation and 
redshift measurement was obtained in the last year.
The \reflex\ II sample as a whole extends the flux limit of \reflex\ I from 
3$\times10^{-12}$ erg/s/cm$^2$ to 1.8$\times10^{-12}$ erg/s/cm$^2$ 
in the 0.1-2.4 keV ROSAT band. This flux limit was imposed on a 
fiducial flux calculation assuming cluster
parameters of 5 keV for the ICM temperature, a metallicity
of 0.3 $Z_{\odot}$, a redshift of $z$=0, and an interstellar
hydrogen column-density according to the 21cm measurements of
Dickey and Lockman (1990). This fiducial flux was 
calculated independent of (prior to) any redshift information
and is therefore somewhat analogous to an extinction-corrected
magnitude limit without K-correction in optical astronomy.
We then calculated the true fluxes and luminosities by
taking an estimated ICM temperature from the X-ray 
luminosity-temperature scaling relation 
(Pratt et al. 2009) and redshifted spectra into account.

\begin{figure}
\begin{center}
\resizebox{\hsize}{!}{\includegraphics[height=6cm]{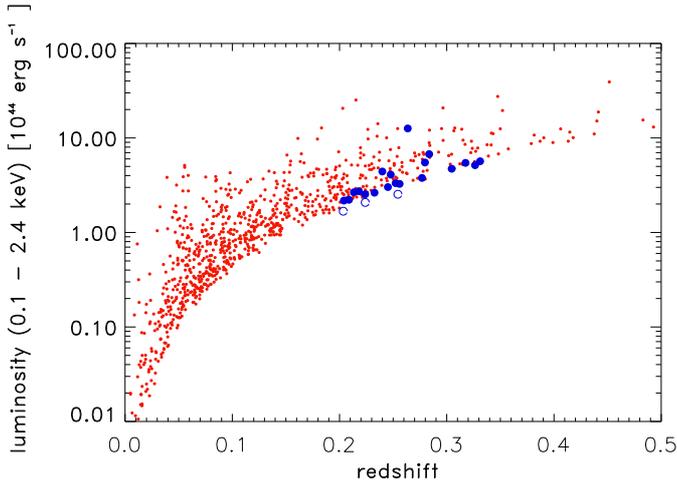}}
\end{center}
\caption{Luminosity-redshift plot of the \reflex\ II 
cluster catalogue. The limiting flux is 1.8 $\times10^{-12}$ 
erg/s/cm$^2$ in the 0.1-2.4 keV ROSAT band. The blue filled 
circles are the 19 clusters in the \reflex\ II catalogue, and 
the 3 open circles below the flux limit are the clusters 
in the \reflex\ III supplementary catalogue with $z>0.2$.}
\end{figure}
\vspace{0.2cm}

Fluxes were measured by adopting a growth curve analysis in an 
aperture maximising the signal. The flux values were then 
corrected to an aperture of $r_{500}$ \footnote{Defined as the
radius where the mean cluster mass density is 500 times
the critical density of the Universe at the epoch
at which the cluster is observed.} by estimating mass 
based on the luminosity-mass relation of Pratt et al.(2009)
in an iterative way.

The total count rates, from which fluxes and luminosities
were determined, were derived with the growth curve analysis (GCA)
as described in B\"ohringer et al. (2000). The cluster
candidates were compiled from a flux-limited sample of
all sources in the RASS analysed by the GCA method,
combining all information on the X-ray detection parameters,
visual inspection of available digital
sky survey images, properties in the NASA extragalactic
database, \footnote{The NASA/IPAC Extragalactic Database (NED) 
is operated by the Jet Propulsion Laboratory, California 
Institute of Technology, under contract with the National 
Aeronautics and Space Administration.} and other available
images at optical or X-ray wavelength. 
In addition, we cross-correlated our data with publicly available
SZ catalogues from large surveys such as {\it Planck}, SPT,
and ACT. For a detailed description 
of the construction of the \reflex\ II galaxy cluster catalogue, 
we refer to B\"ohringer et al. (in prep.).

The current cluster sample, comprising 19 objects,
was selected from among the clusters of the \reflex\ II
sample confirmed by means of follow-up in 2010-2011 at 
redshifts above $z \ge 0.2$, and the remaining three objects
were confirmed from the data taken in 2004. These clusters 
have luminosities in the range $L_{X}$=1.6--13$\times 10^{44}$erg/s 
and estimated masses (based on Pratt et al. 2009) of $M_{500}$=2.3--7.8
$\times 10^{14}$ $M_{\odot}$, and cover 
the redshift range $z$=0.2--0.33. 
Fig. 1 shows the X-ray luminosity and redshift distribution
of the \reflex\ II cluster sample and the newly confirmed 19 clusters
and 3 supplementary clusters in comparison. The newly added cluster
which has an exceptionally high luminosity is RXCJ1914.5-5928.
Table 1 lists optical
properties, and the essential X-ray properties
of the galaxy clusters are found in Table 2. For 
the derivation of distance-dependent parameters,
 we use a flat cosmology with 
$\Omega_m$=0.3, $h_{70}$=$H_{0}$/70 km/s/Mpc.

\section[]{Optical spectroscopic data}

\subsection[]{Observations}

A more accurate determination of physical parameters is possible 
for spectroscopically confirmed clusters than for those with
photometric redshifts, which usually have larger errors. 
In addition, a well-observed 
redshift sample of cluster members probes the dynamical state of 
a cluster that can be studied with the galaxy velocity 
dispersion within the system.

With this in mind, \reflex\ and some NORAS clusters have been
spectroscopically followed-up since 1992 with the EFOSC1/2
instruments.
The 19 cluster data presented in this paper is based on two 
spectroscopic observing campaigns, 085.A-0730(A) and 086.A-0055(A), 
during five nights each in September 2010 and March 2011 with the EFOSC2 
instrument at the New Technology Telescope (NTT) in La Silla.
The additional three clusters are from observations in 2004 with 
the EFOSC2 instrument at the ESO 3.6m telescope in La Silla.

Where possible, the multi-object spectroscopy (MOS) mode of 
EFOSC2 was used, and in some cases if a visual inspection of 
the pre-imaging data identified a clear BCG, we used a single long-slit (LS) 
observation on the BCG with at least one other member of the cluster. 
The different observing modes are marked in Table 1, 
which also lists the redshifts from the spectroscopic observations.

Each field of our candidate clusters was optically imaged in 
Gunn r band around the X-ray centre for the target selection 
and the mask making.
The imaging resolution is 0.12 x 0.12 arcsec$^2$, and the field 
of view is 4.1 x 4.1 arcmin$^2$ for both imaging and spectroscopic 
observations. When necessary, the field was rotated to help optimise
target selection. We used the grism that covers the wavelength 
range between 4085 $\AA$ and 7520 $\AA$ with 1.68 $\AA$ per pixel 
at resolution 13.65 $\AA$ per arcsec. We used 
slitlets with a fixed width of 1.5 arcsec for the MOS and of 
2.0 arcsec for the long-slit observations. 

Including at least three bright objects, preferably stars, 
to orient the field, the slitlets were allocated to the 
candidate member galaxies. Typically ten slitlets per field were 
punched onto one mask. Owing to the instrumental set-up, a maximum of 
five MOS plates can be loaded per night. The typical exposure
time for the clusters in this paper ranges from 1500 to 2000 
seconds per exposure, whose variation also reflects the 
observing conditions. The seeing was on average around one 
arcsec for most of the observations.

\subsection[]{Data reduction}

We reduced the data using the standard reduction pipeline of 
IRAF. The reduction scheme is briefly summarised below, 
which applies equally to MOS and single-slit
data.

{\bf Bias correction} The CCD frames were visually inspected 
and trimmed at all edges of the images. All bias frames of a
similar mean in each run were combined to reduce the rms 
fluctuations in the master bias, which was then subtracted from the 
science images. 

{\bf Flatfielding} Unlike the bias that is an additive correction,
the flatfielding corrects the multiplicative spatial variations. 
The correction includes the varying quantum efficiency manifested 
as the structural variations in the CCD, and vignetting.
We combined all dome flats for the combination of our slit, 
grism, and filter as described in the previous subsection. 
The spectrum of the quartz lamp was then fitted with a cubic 
spline function, and normalised. All spectroscopic exposures
were then divided by the resulting response frame.

{\bf Extracting the spectra}
After the bias and flatfield corrections, the two science spectra were 
combined and weighted using IMCOMBINE, which also removes 
the cosmic-ray events. We used APALL in IRAF to extract the spectra.
The one-dimensional collapsed spectra were sky-background subtracted by using
the sky region on both sides of an object spectrum as a background 
estimate.
The calibration spectra, which are two He-Ar frames were also 
extracted from the same spatial position as the science spectra.

{\bf Wavelength calibration}
We identified at least 4-5 known lines in the He-Ar lamp spectrum
to fit the wavelength as a function of pixel number with an
rms of far smaller than 1 $\AA$ where the corresponding rms of the
long-slit observation is larger owing to the larger slit 
width. The procedure was performed with the IDENTIFY and REIDENTIFY
packages. 

{\bf Astrometric correction}
The optical image taken during the spectral observation 
does not contain sky coordinates. The first guess was taken 
from the pointing directions used for the telescope, and 
the image was then astrometrically corrected for based on
the accurate location of a number of identified objects
in a public optical catalogue.

\begin{figure}
\begin{center}
\resizebox{\hsize}{!}{
\includegraphics[height=6cm]{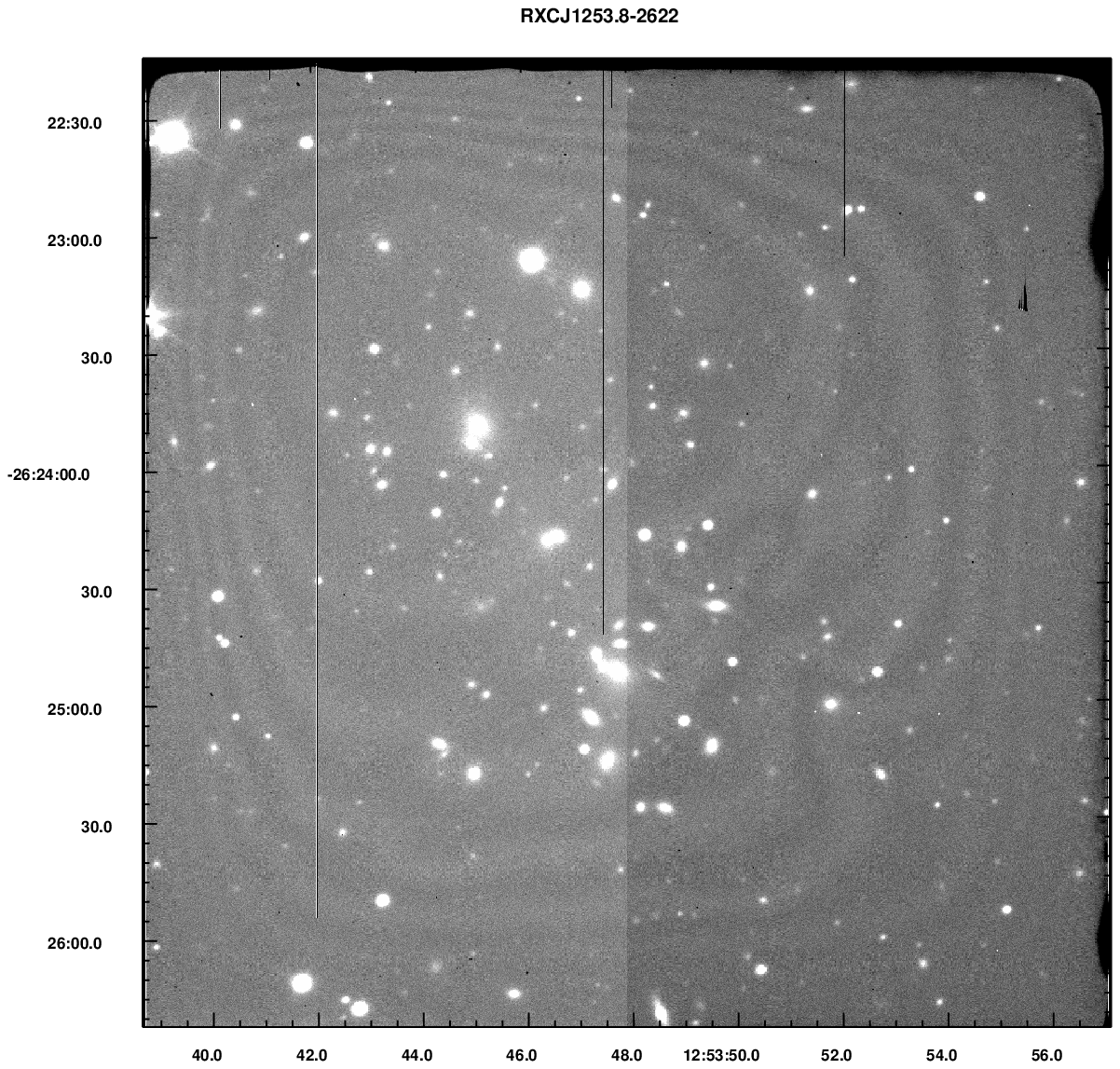}
}
\resizebox{\hsize}{!}{
\includegraphics[height=6cm]{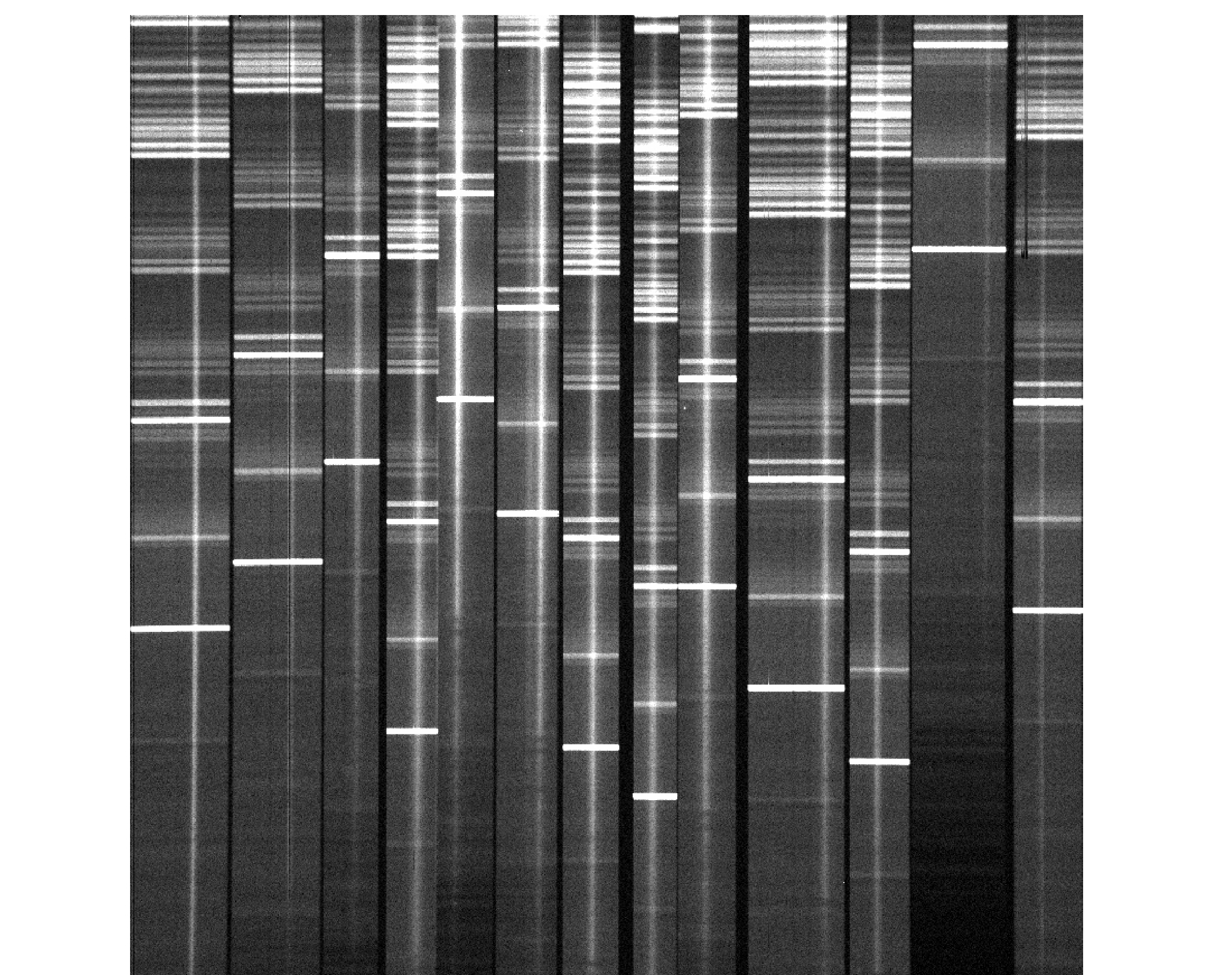}
}
\resizebox{\hsize}{!}{
\includegraphics[height=6cm,angle=-90]{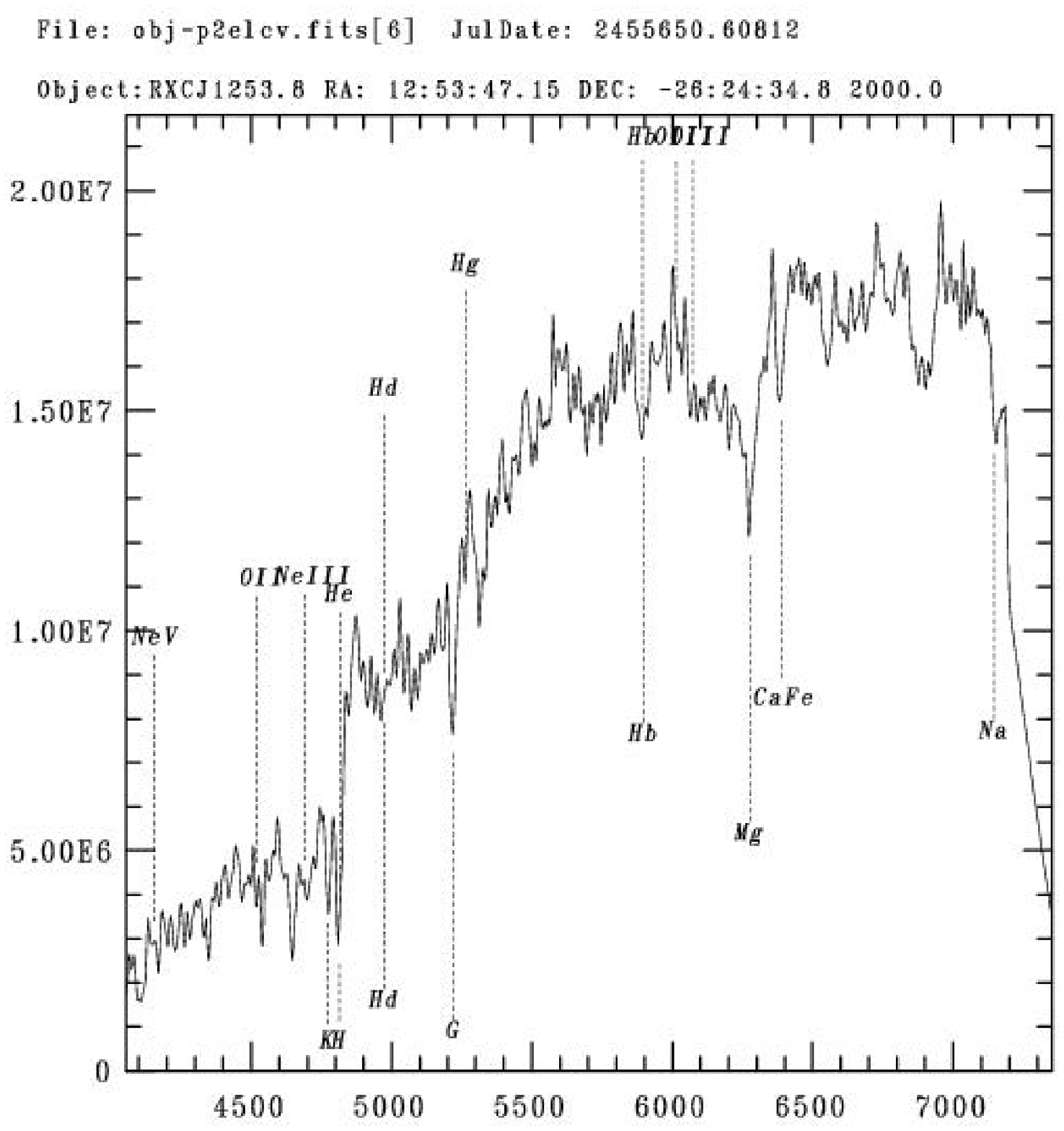}
}
\end{center}
\caption{The optical image (top), cleaned MOS frame (middle), and an 
example spectrum (bottom) of RXCJ1253.8-2622. The bottom panel 
shows the extracted spectrum with the identified absorption 
lines from the 6th object from the left
of the middle panel.}
\end{figure}
\vspace{0.2cm}

\subsection[]{Redshift determination}

The calibrated spectra undergo several corrections such as 
skyline removals, a heliocentric correction, and emission
line removals before the redshift determination. The redshifts
from the emission lines are determined separately after 
the correlation with the passive galaxy templates, hence 
we removed them for this step. We then use the RVSAO package, 
which applies the cross-correlation technique to the input 
templates of galaxy spectra to measure the object redshift. 
The \reflex\ I templates were used for this analysis, which 
include 17 galaxy and stellar templates (Guzzo et al. 2009).
In essence, the method uses the fast fourier transform to
construct the correlation function of the input spectrum 
together with the template, after filtering to remove
the spurious components and binning noise. The significance
of the correlation is reflected in the amplitude of the correlation
peaks, and the shift in the correlation function corresponds
to the velocity of the measured object. Among all correlation
functions, we selected the one with the highest R-value as the
most closely matching spectrum, and quoted accordingly its redshift 
measurement. We confirmed a spectroscopic cluster detection if at 
least three galaxies have their R value greater than 5, and lie 
within $\pm$3000 km/s of the mean velocity of the cluster members.
We then took the median of those galaxy redshifts as 
the cluster redshift.

For the long slit observations, a cluster was confirmed if the 
redshift of the BCG was known and at least another galaxy has
a similar redshift within the aforementioned criteria.

A small fraction of our cluster galaxy sample exhibit emission lines,
which are identified with EMSAO in the RVSAO package. These
are marked by E in Tables A.1--3. The resulting individual galaxy
redshifts for the clusters in the \reflex\ II sample are listed
in the Tables A.1--3 in the appendix. 

\section[]{Properties of the cluster sample}

In Table 2, we list the X-ray properties of our cluster sample.
The X-ray flux and luminosity in the 0.1-2.4 keV band are calculated
iteratively from the count rate using an estimated temperature
and the measured redshift and $N_{H}$ for the count rate to flux conversion and
the luminosity K-correction to yield the rest-frame luminosity. We use the 
luminosity-temperature relation from Pratt et al.(2009), $T_{X}$ = 
3.306~$L_{X}^{1/3} h_{70}^{2/3}$, for the temperature estimate, where 
$T_{X}$ is in units of keV and $L_{X}$ in units of $10^{44}$ erg/s.
This flux and luminosity were measured for the largest aperture used
by the GCA method. The size of this aperture is also given in Table
2 in units of arcmin. 
The uncertainties in the flux and luminosity
measurements were calculated from the contributions of the Poisson statistics
of the source counts and the variance in the background determined
in a total area of 4009 arcmin$^2$. The growth curve
result was carefully checked for blending of the target source with
neighbouring sources and a manual deblending was performed when necessary.
The additional flux error introduced by the deblending was estimated
and added to the above statistical error. A complex case of deblending
is discussed below and shown in Fig. 3.

\begin{figure}
\begin{center}
\resizebox{\hsize}{!}{
\includegraphics[height=4cm]{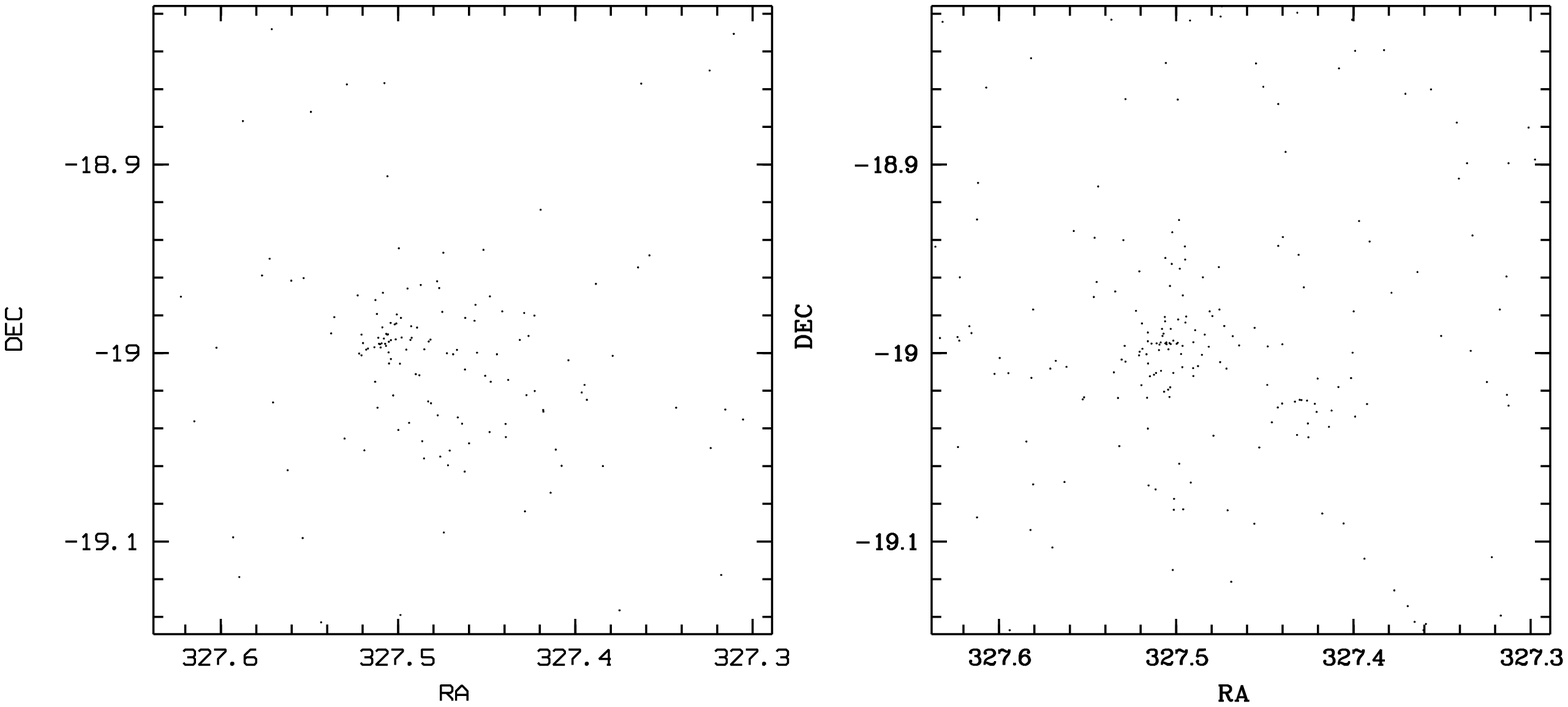}
}
\resizebox{\hsize}{!}{
\includegraphics[height=3.8cm]{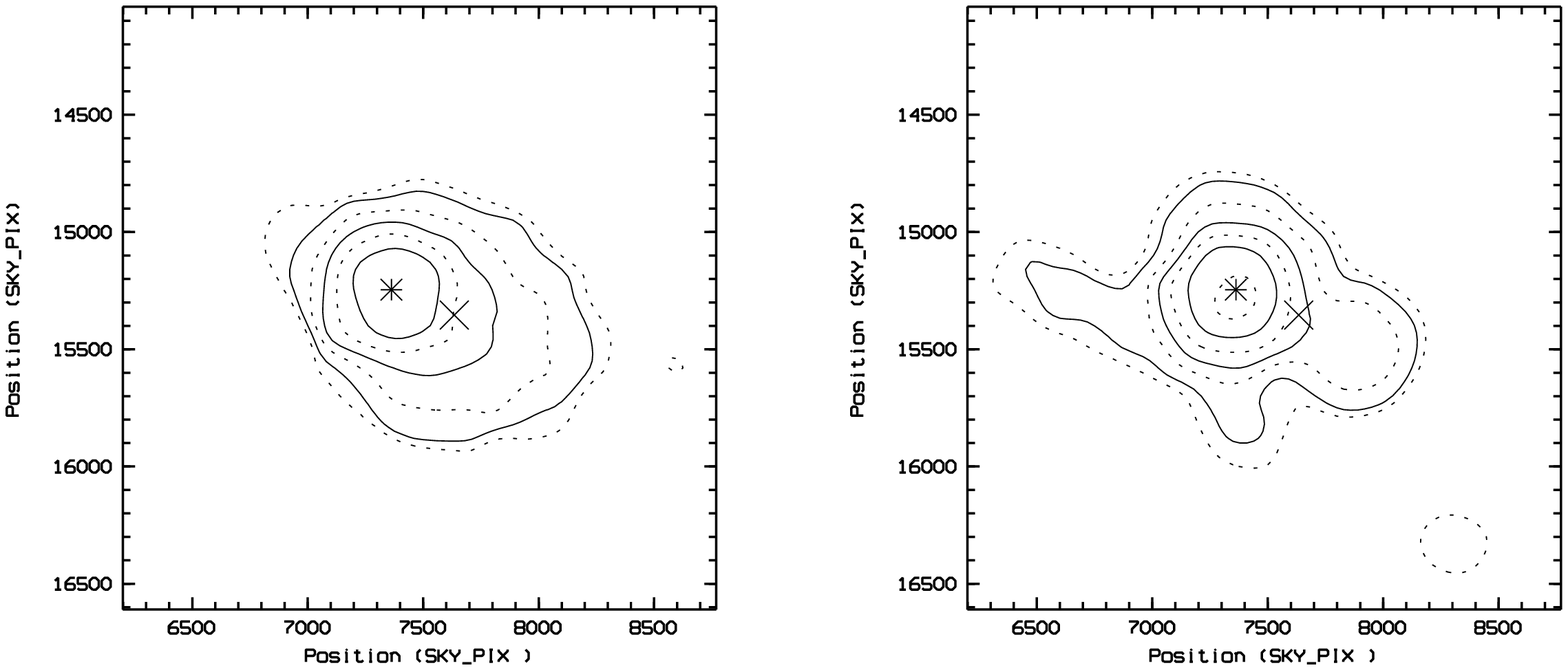}
}
\end{center}
\caption{RXCJ2149.9-1859 : (upper panel) photon distribution 
in the hard band (left) and soft band (right). 
Most of the emission in the soft band is identified as the Seyfert 
galaxy occupying a region in the north-east of the cluster centre.
(lower panel) Contour plots of the significance in flux 
in the hard (left) and soft band (right) at 1.5, 2, 3, 4, 5, 6, 
7$\sigma$ per Gaussian beam of 1 arcmin width. 
Contours are marked with two alternating types of lines for clarity.
After deblending, the centre of the cluster moves to where the cross is.
Note that the location of the Seyfert galaxy is marked by an asterisk.
}
\end{figure}
\vspace{0.2cm}

We also corrected the aperture luminosity to
the luminosity inside $r_{500}$ for which we used the $L_{X} - M_{Y}$ relation
of Pratt et al.(2009), which translates to $r_{500}$=0.895~Mpc $L_{X}^{0.195}
{E(z)}^{-1.056}$, with $L_X$ in units of $10^{44}$ erg/s for the 
0.1 to 2.4 keV band. Further details of these calculations will be
described in a forthcoming paper on the \reflex\ II sample construction
by B\"ohringer et al.

\subsection[]{Detection at other wavelengths}

Some of the clusters in our list are sufficiently prominent that 
seven were previously detected by Abell (1958, 1989) and one was noted 
by Zwicky, as indicated in Table. 1. One of the clusters, RXCJ1914.5-5928, 
has been detected by the {\it Planck} survey (Planck Collaboration 2011a).

\subsection[]{Additional diagnostic X-ray properties}

Given the low count rates (19-65 counts and two cases
with 10 and 16 counts), the low angular resolution of the RASS
(90 arcsec half power radius), and 
the low spectral resolution, the data do not permit a 
detailed spectral and morphological analysis to be made for
these sources. However, it allows us to derive two simple and 
robust parameters.

One parameter, the spectral hardness ratio, is determined from
the count rates in the soft (channel 11-40, $\sim$ 0.1-0.4 keV)
and hard (channel 52-201, $\sim$ 0.5-2 keV) bands by
the formula HR=(H-S)/(H+S). Since for a given temperature, redshift, 
and assumed metal abundance of 0.3 solar the expected X-ray spectrum
of a cluster is known, the hardness ratio for a given interstellar
absorption can be calculated. In Fig. 4, we show the distribution 
of the deviations of the detected from the estimated values of
the hardness ratio in units of significance for the total \reflex\ II
catalogue, the present cluster sample, and we compare these properties
to those of a sample of non-cluster sources.
\footnote {These non-cluster sources comprise 
1524 of the RASS X-ray sources in the \reflex\ sky region with
fluxes above the \reflex\ II cut that were described as cluster
candidates in the \reflex\ II sample construction 
(B\"ohringer et al. 2001).} Fewer than one third of the 
non-cluster sources overlap with the cluster distribution.
Therefore, the inspection of the hardness ratios is a powerful
diagnostic of the possible AGN (or other non-cluster X-ray sources)
contamination of our sample. Indeed all clusters of the present
sample deviate by less than 3$\sigma$ from expectations,
except for the cluster RXCJ2149.9-1859, where the contamination 
is obvious and has been deblended as described below.

\begin{figure}
\begin{center}
\resizebox{\hsize}{!}{
\includegraphics[height=6cm]{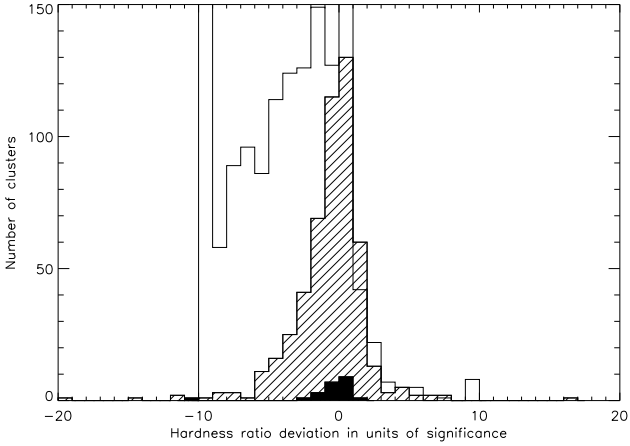}
}
\end{center}
\caption{Distribution of the measured hardness ratio 
deviations from the expectation in units of the significance 
for the \reflex\ II 
clusters is marked with the slanted lines in the histogram. 
The filled bars show the distribution of the cluster 
sample in this paper. 
The blank bars mark 1542 non-cluster 
sources identified during the construction 
of the \reflex\ II catalogue. It is clear that the hardness
deviation of the clusters has an approximate Gaussian distribution 
with a mean close to zero, while the non-cluster sources have 
a significant excess towards the negative deviations.}
\end{figure}
\vspace{0.2cm}

The second diagnostic is used to discriminate extended
sources from either point or unresolved sources. The most 
robust test we have applied in the \reflex\ project is a Kolmogorov-Smirnov
test to check if the source photon distribution was consistent with 
a point source and the estimated background. The test was conducted
in a radial region of 6 arcmin radius around the cluster centre
(B\"ohringer et al. 2000). We assumed a probability threshold
of $<1\%$ for a reliable designation of a source as extended.
We found a significant extent for 11/22 
of the clusters in our sample, which is again a very good 
confirmation given that the clusters have a high
redshift and are not always expected to display a significant extent.
In comparison, the \reflex\ I sample at higher flux levels show
a significant extent for 81 \% of the clusters with unresolved
sources mostly at higher redshifts (Fig. 26 in B\"ohringer et al.
2001).
Fig. 5 shows the distribution of the extent parameter,
-$log$(KS probability) for the \reflex\ II catalogue, the present
sample, and the 1542 non-cluster sources. Only a few percent of the
non-cluster sources have an extent parameter larger than 2.
They are mostly nearby galaxies, very bright stars with an
artificial extent signal, and some sources with spurious extents.
Fig. 6 shows the hardness ratio - extent parameter distribution
for all \reflex\ II clusters and the present sample. Most of 
the present clusters with an extent parameter smaller than 2,
which is used as a conservative threshold, still have parameter 
values corresponding to a low probability of being a point source.

\begin{figure}
\begin{center}
\resizebox{\hsize}{!}{
\includegraphics[height=6cm]{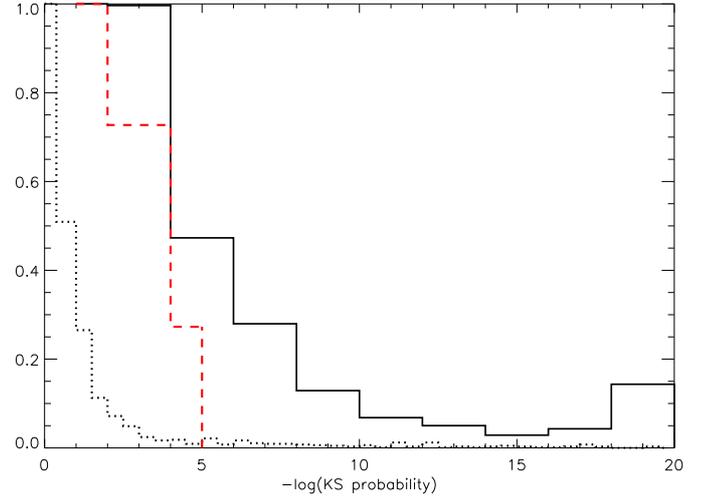}
}
\end{center}
\caption{Fractional histograms of the -$log$ Kolmogorov-Smirnov 
probability for three categories of objects. The solid line 
represents the \reflex\ II clusters as in Fig. 4. We truncate
the values so that the last bin contains the remaining counts for
the whole sample, which amounts to 5 \%. 
The dotted lines mark 1542 non-cluster sources, 
and the red dashed lines are for the 22 clusters.
The non-cluster sources are strongly concentrated
at small extents.}
\end{figure}
\vspace{0.2cm}

\begin{figure}
\begin{center}
\resizebox{\hsize}{!}{
\includegraphics[height=6cm]{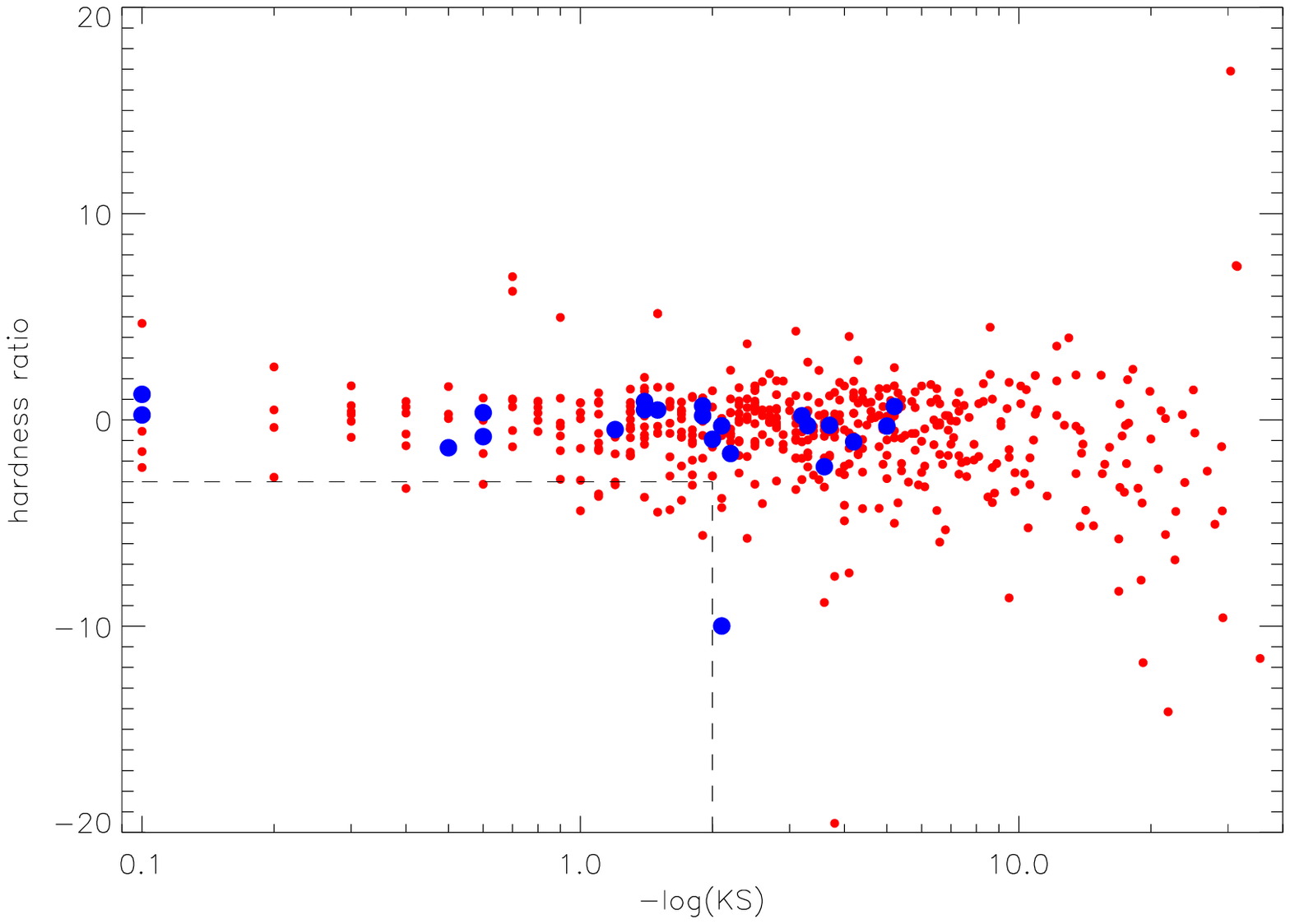}
}
\end{center}
\caption{Hardness ratio vs. -$log$ probability of the
Kolmogorov-Smirnov test for the same clusters in the Figure 2.
The larger blue circles indicate the 22 clusters presented in this
paper. The lower left sector enclosed by dashed lines is 
the zone where there is a high chance that a source is not
a cluster. We note that this is a conservative limit, and
any source that falls into this category must be scrutinised 
in detail to be confirmed. 
One cluster indicated by a blue circle that has an exceptionally soft 
emission among all of the 22 (the large blue circle nearest to the vertical 
dashed line) is RXCJ2149.9-1859. See section 4.3 for 
detailed remarks about this source. Two clusters of small extent
are clearly compact clusters where there is a central dominant galaxy 
close to the X-ray centres.
}
\end{figure}
\vspace{0.2cm}

\subsection[]{Remarks on individual clusters}

\indent \indent RXCJ0347.4-2129 has a complex emission region with two
major components.

RXCJ0521.4-2754 is deblended from a nearby source.

RXCJ1253.8-26 has a complex, diffuse emission region.
The structure of this source will be inspected in greater detail using
an allocated 7ks Chandra observation.

The centre of RXCJ1914.5-5928 coincides with that of PLCKG337.1-26.0 
to within a distance of 4.3 arcmin from the SZ cluster center 
(Planck Collaboration 2011a). 
In terms of the flux limit this cluster should have been included 
in the REFLEX I catalogue. However, the REFLEX I catalogue was 
strictly based on the detection of X-ray sources, as well as 
significant galaxy overdensities, in the COSMOS database, which is
a relatively shallow with a non-uniform coverage across the sky
(B\"ohringer et al. 2001). Hence, this cluster was identified 
as a supplementary cluster on the basis of its X-ray extent. 
With a more comprehensive identification procedure for the 
REFLEX II catalogue, it is identified as a part of the new sample.

The cluster, RXCJ2149.9-1859, contains an X-ray point source that
is offset by about 2.4 arcmin from the X-ray centre of the cluster. 
The point source
is significantly softer than the cluster emission and originates 
from a Seyfert galaxy at $z$=0.157. 
It is located at 21:49:58 in RA and -18:59:24 in Dec and
is denoted by an asterisk in the lower panels of Fig. 3.
The flux from this source was deblended from the cluster emission.
The X-ray surface brightness distribution of this source
in the soft (0.1-0.4 keV) and hard band (0.5-2 keV) is 
shown in Fig. 3. The contribution of the Seyfert galaxy 
to the cluster X-ray emission is clearly noticeable 
as a much softer source than the cluster emission north-east of 
the cluster centre. It was deblended
by cutting off the contaminated sector of the cluster emission
and filling it with the remaining azimuthal average. 
After deblending, the new centre is then found to be 
1.5 arcmin south of its initial location.

RXCJ2149.0-3228 has been deblended from a source in the 
south-west of the cluster. 

\begin{table*}
\begin{center}
\caption{Optical properties of all clusters studied}
\label{table1}
\centering
\begin{tabular}{c c c c c c c c}
\hline
\hline
\multicolumn{1}{c}{Cluster} & 
\multicolumn{1}{c}{Alt. Name} & 
\multicolumn{1}{c}{R.A.} & 
\multicolumn{1}{c}{Dec.} & 
\multicolumn{1}{c}{$z$} & 
\multicolumn{1}{c}{$\Delta z$} & 
\multicolumn{1}{c}{$N_{gals}$} & 
\multicolumn{1}{c}{Observing mode}\\
(1) & (2) & (3) & (4) & (5) & (6) & (7) & (8)\\
\hline
RXCJ0052.4-5746 &      & 13.1093  & -57.7806 & 0.2559 & 2.37 & 3 & MOS \\
RXCJ0129.4-6432 &      & 22.3730  & -64.5409 & 0.3264 & 1.98 & 6 & MOS \\
RXCJ0139.5-2629 &      & 24.8888  & -26.4886 & 0.2445 & 1.56 & 4 B & LS \\
RXCJ0343.5-7152 &      & 55.8775  & -71.8764 & 0.2136 & 1.61 & 3 & MOS \\
RXCJ0347.4-2149 & A3168& 56.8724  & -21.8224 & 0.2399 & 1.77 & 5 & MOS \\
RXCJ0425.5-0045 & ZwCl0423.2-0053 &  66.3798 & -0.7577  & 0.2240 & 1.75 & 6 & MOS \\
RXCJ0521.4-2754 &      & 80.3581  & -27.9065 & 0.3175 & 2.79 & 2 B & LS \\
RXCJ1253.8-2622 & A1633& 193.4682 & -26.3828 & 0.2179 & 1.65 & 9 B & MOS \\
RXCJ1617.6-0714 &      & 244.4099 & -7.2446  & 0.2523 & 2.36 & 3 B & MOS \\
RXCJ1914.5-5928 & PLCKG337.1-26.0 & 288.6333 & -59.4767 & 0.2636 & 1.74 & 4 & MOS \\
RXCJ2006.4-1912 &      & 301.6046 &-19.2106  & 0.2769 & 1.83 & 4 & MOS \\
RXCJ2041.2-1324 &      & 310.3188 & -13.4149 & 0.3311 & 2.11 & 5 B & MOS \\
RXCJ2049.9-3029 & A3715& 312.4885 & -30.4926 & 0.2088 & 1.55 & 3 & MOS \\
RXCJ2126.1-3202 & AS0946& 321.5441& -32.0338 & 0.2797 & 1.66 & 2 B & MOS \\
RXCJ2149.0-3228 &      & 327.2524 & -32.4744 & 0.2326 & 1.94 & 4 & MOS \\
RXCJ2149.9-1859 &      & 327.4500 & -19.0050 & 0.2836 & 3.64 & 2 B & LS\\
RXCJ2308.3-0155 &      & 347.0873 & -1.9210  & 0.3047 & 2.07 & 4 & MOS \\
RXCJ2317.9-3534 & A3989& 349.4828 & -35.5794 & 0.2041 & 2.04 & 3 & MOS \\
RXCJ2351.0-1954 &      & 357.7703 & -19.9132 & 0.2477 & 1.90 & 2 & MOS \\
\hline
RXCJ1013.1-1641 & & 153.2794 & -16.6884 & 0.2240 & 2.00 & 3 & LS \\
RXCJ1023.2-0636 & A1001& 155.8268 & -6.6046 & 0.2544 & 1.06 & 3 & LS \\
RXCJ1355.6-2623 & A1816& 208.9334 & -26.3833 & 0.2035 & 1.43 & 5 & MOS \\
\hline
\end{tabular}
\end{center}
Notes: (1) R.A. and Dec. for J2000 in degrees. 
(2) The centre of RXCJ1914.5-5928 is 4.3 arcmin away from
that of PLCKG337.1-26.0. There are three different redshifts
associated with this {\it Planck} cluster (Planck collaboration 
2011a, 2011b). We take the redshift of the component with the largest 
contribution to the Fe line at 0.26 in the Table 2 of Planck 
Collaboration 2011b.
(6) Errors in units of $10^{-4}$
(7) Number of galaxies with concurrent redshifts, B indicates when the 
BCG redshift is available. LS marks the redshift obtained from a single 
long-slit observation. The last three entries are based on the observation 
in March 2004 with the EFOSC2 instrument, which forms part of the 
\reflex\ III supplementary catalogue. 
\end{table*}
\begin{table*}
\begin{center}
\caption{X-ray properties of the sample clusters}
\label{table2}
\centering
\begin{tabular}{c c c c c c c c c c c}
\hline
\hline
\multicolumn{1}{c}{Cluster} & 
\multicolumn{1}{c}{Count Rate} & 
\multicolumn{1}{c}{$F_X$} & 
\multicolumn{1}{c}{$L_X$} & 
\multicolumn{1}{c}{$L_{X,r_{500}}$} & 
\multicolumn{1}{c}{Error} &
\multicolumn{1}{c}{$r_{500}$} &
\multicolumn{1}{c}{Ext} &
\multicolumn{1}{c}{$N_H$} &
\multicolumn{1}{c}{HR} &
\multicolumn{1}{c}{KS prob} \\
(1) & (2) & (3) & (4) & (5) & (6) & (7) & (8) & (9) & (10) & (11)\\
\hline
RXCJ0052.4-5746 & 0.090 & 1.86&  3.33&  3.26& 28.9 & 0.988&  6.5& 3.40 &0.96 &1.4\\
RXCJ0129.4-6432 & 0.090 & 1.80&  5.46&  5.14& 42.1 & 1.043&  8.0& 2.24 &0.50 &3.7\\
RXCJ0139.5-2629 & 0.069 & 1.96&  3.19&  3.02& 30.4 & 0.985&  8.5& 1.49 &0.24 &2.0\\
RXCJ0343.5-7152 & 0.108 & 2.34&  2.82&  2.66& 20.4 & 0.980&  9.5& 5.31 &0.76 &3.3\\
RXCJ0347.4-2149 & 0.153 & 3.11&  4.80&  4.41& 20.3 & 1.070& 13.0& 2.83 &0.59 &5.0\\
RXCJ0425.5-0045 & 0.087 & 1.94&  2.60&  2.53& 24.1 & 0.959&  7.0& 6.38 &0.32 &3.6\\
RXCJ0521.4-2754 & 0.100 & 1.98&  5.62&  5.45& 29.5 & 1.055&  6.5& 1.77 &0.66 &1.9\\
RXCJ1253.8-2622 & 0.100 & 2.25&  2.84&  2.73& 58.8 & 0.979&  8.0& 6.80 &1.00 &1.9\\
RXCJ1617.6-0714 & 0.073 & 1.89&  3.28&  3.32& 26.0 & 0.987&  5.5& 12.45&0.63 &2.2\\
RXCJ1914.5-5928 & 0.320 & 7.07& 13.16& 12.60& 20.0 & 1.285&  9.0& 5.82 &0.68 &4.2\\
RXCJ2006.4-1912 & 0.076 & 1.73&  3.68&  3.58& 52.6 & 0.995&  6.5& 7.22 &1.00 &0.6\\
RXCJ2041.2-1324 & 0.093 & 1.94&  6.05&  5.66& 18.3 & 1.061&  8.5& 3.65 &0.86 &1.4\\
RXCJ2049.9-3029 & 0.087 & 1.92&  2.23&  2.21& 38.0 & 0.939&  6.5& 6.16 &0.80 &2.1\\
RXCJ2126.1-3202 & 0.099 & 2.11&  4.56&  5.49& 25.0 & 1.036&  3.5& 4.54 &1.00 &0.1\\
RXCJ2149.0-3228 & 0.075 & 1.49&  2.19&  2.14& 36.0 & 0.922&  6.5& 2.09 &0.28 &0.5\\
RXCJ2149.9-1859 & 0.151 & 3.12&  6.94&  6.74& 22.0 & 1.121&  7.0& 3.35 &-0.04&2.1\\
RXCJ2308.3-0155 & 0.089 & 1.88&  4.91&  4.72& 22.5 & 1.035&  7.0& 4.29 &1.00 &5.2\\
RXCJ2317.9-3534 & 0.101 & 1.98&  2.19&  2.17& 59.4 & 0.938&  6.5& 1.69 &0.18 &0.6\\
RXCJ2351.0-1954 & 0.133 & 2.61&  4.33&  4.09& 19.5 & 1.045&  9.0& 1.50 &0.50 &3.2\\
\hline
RXCJ1013.1-1641 & 0.071 & 1.54&  2.07&  1.95& 18.3 & 0.917 & 9.0& 5.36 &0.90 &0.1\\
RXCJ1023.2-0636 & 0.067 & 1.43&  2.53&  2.84& 19.4 & 0.938 & 4.0& 4.65 &0.94 &1.5\\
RXCJ1355.6-2623 & 0.071 & 1.53&  1.67&  1.63& 43.7 & 0.890 & 7.0& 5.02 &0.69 &1.2\\
\hline
\end{tabular}
\end{center}
Notes: (2) in s$^{-1}$ (3) in units of $10^{-12}$ erg/s/cm$^2$
in [0.1-2.4] keV band measured in the aperture column
(4) $L_{X}$ in $10^{44}$ erg/s in [0.1-2.4] keV band in the 
aperture
(5) $L_{X}$ in $10^{44}$ erg/s in [0.1-2.4] keV band measured 
in $r_{500}$ (6) error in percentage 
(7) $r_{500}$ in Mpc (8) source extent in arcmin
(9) $N_{H}$ in units of $10^{-20}$ cm$^{-2}$ 
(10) hardness ratio defined as HR=(H-S)/(H+S) 
where H is the count rate in the hard band (0.5-2 keV) and S 
in the soft band (0.1-0.4 keV) (11) -$log$ probability of
the Kolmogorov-Smirnov test.
\end{table*}

\section[]{Summary and conclusions}

We have presented a sample of 22 nearby spectroscopically confirmed 
clusters selected from the \reflex\ II catalogue. These clusters
all have redshifts above $z$=0.2 and therefore for the given
flux limit of the sample they are very X-ray luminous and 
have high estimated masses. This implies that these objects are 
interesting for follow-up studies, in particular for
Sunyaev-Zel'dovich effect observations, gravitational lensing
measurements, studies of the dark matter and baryon mass fraction, 
and cosmological studies. For this reason, we wish to make these 
findings public in advance of the publication of the main catalogue.
For the two most prominent clusters of the sample, RXCJ1914.5-5928 
and RXCJ2149.9-1859 we have obtained mass estimates, $M_{500}$, 
of 7.8$\times 10^{14}M_{\odot}$ and 5.3$\times 10^{14}M_{\odot}$ using the 
scaling relation of Pratt et al. (2009), 
$M_{500}$=2.02 $L_x^{0.585}$ $E(z)^{-1.17}$. 
The most luminous cluster, RXCJ1914.5-5928, was detected among
the high significance sample in the {\it Planck} survey.
These two most massive clusters are the most interesting objects, 
e.g. for lensing studies and with estimated $Y_{X}$ parameters of 
8.6$\times 10^{14}M_{\odot}$keV and 4.4$\times 10^{14}M_{\odot}$keV 
(using the $Y_{X}$ scaling from Arnaud et al. 2007, 
$Y_x$ = 0.1934 $M_{500}^{1.825}$ $E(z)^{0.73}$),
they should be detectable by on-going Sunyaev-Zel'dovich experiments.

\begin{acknowledgements} 
We thank the anonymous referee for helpful comments.
\end{acknowledgements}
%


\appendix
\section{List of spectroscopic redshifts of the identified cluster galaxies}

\begin{longtable}{c c c c c}
\caption{List of the cluster galaxies : Notes: E in (5) 
denotes a galaxy with emission lines. The first galaxy of 
RXCJ2126.1-3202 also has emission lines whose redshift is 
provided adjacent to E in brackets. We take also the literature 
galaxy redshift in the vicinity of the cluster centres, 
available in the NED search, which are marked by its object 
identifiers in the catalogues. (C) refers to the galaxies 
found in Colless et al.(2003), and (J) in Jones et al.(2009).} \\
\hline
\hline
Cluster & R.A. & Dec. & z & remark\\
(1) & (2) & (3) & (4) & (5)\\
\hline
\endfirsthead
\caption{continued.}\\
\hline
\hline
\endhead
\hline
\endfoot
RXCJ0052.4-5746 & & & \\
                & 13.133868 & -57.775865 & 0.2493 &  \\
		& 13.147719 & -57.776764 & 0.2559 & E\\
		& 13.170916 & -57.794    & 0.2582 &  \\
RXCJ0129.4-6432 & & & \\
                & 22.435034 & -64.597408 & 0.3204 \\
		& 22.437597 & -64.577611 & 0.3264 \\
		& 22.430907 & -64.570233 & 0.3290 \\
		& 22.438965 & -64.554441 & 0.3246 \\
		& 22.459002 & -64.549708 & 0.3287 \\
		& 22.429443 & -64.545141 & 0.3243 \\
RXCJ0139.5-2629 & & & \\
                & 24.868537 & -26.495456 & 0.2451 \\
		& 24.894092 & -26.486702 & 0.2433 \\
		& 24.901974 & -26.48409  & 0.2447 \\
		& 24.901285 & -26.49041  & 0.2451 \\
                 & 24.8275 & -25.4239 & 0.2440 & 2dFGRS S151Z172(C)  \\
                 & 24.8508 & 26.3783 & 0.2462 & 2dFGRS S151Z165(C) \\
                 & 24.7742 & 26.5789 & 0.2506 & 2dFGRS S220Z172(C) \\
RXCJ0343.5-7152 & & & \\
                & 55.945821 & -71.849902 & 0.2147 \\
		& 55.959294 & -71.856585 & 0.2128 \\
		& 56.041557 & -71.889731 & 0.2133 \\
RXCJ0347.4-2149 & & & \\
                & 56.93468  & -21.827332 & 0.2440 \\
		& 56.9313   & -21.8214	 & 0.2399 \\
		& 56.923154 & -21.807932 & 0.2391 \\
		& 56.92696  & -21.803433 & 0.2413 \\
		& 56.914577 & -21.794706 & 0.2397 \\
RXCJ0425.5-0045& & & \\
                & 66.367106 & -0.77938041 & 0.2239 \\
		& 66.371827 & -0.77442204 & 0.2239 \\
		& 66.387509 & -0.75881324 & 0.2178 \\
		& 66.385977 & -0.75223203 & 0.2200 \\
		& 66.40343  & -0.7513067  & 0.2278 \\
		& 66.405553 & -0.74196309 & 0.2242 \\
RXCJ0521.4-2754 & & & \\
                & 80.344202 & -27.90332  & 0.3163 \\
		& 80.356403 & -27.898519 & 0.3187 \\
RXCJ1253.8-2622 & & & \\
                & 193.43474 & -26.41951  & 0.2178 \\
		& 193.43792 & -26.396884 & 0.2159 \\
		& 193.43035 & -26.40101  & 0.2147 \\
		& 193.4442  & -26.404795 & 0.2124 \\
		& 193.44809 & -26.420722 & 0.2234 \\
		& 193.45253 & -26.424123 & 0.2208 \\
		& 193.4567  & -26.409775 & 0.2116 \\
		& 193.46578 & -26.41682  & 0.2234 \\
		& 193.46967 & -26.421824 & 0.2259 \\
                 & 193.448& -25.5856 &  0.2252 & 6dF J1253475-262452(J)\\
RXCJ1617.6-0714 & & & \\
                & 244.40231 & -7.2400582 & 0.2522 \\
		& 244.41683 & -7.2274564 & 0.2487 \\
		& 244.39301 & -7.2549254 & 0.2545 \\		     
RXCJ1914.5-5928 & & & \\
                & 288.59075 & -59.468145 & 0.2624 \\
		& 288.64634 & -59.46307  & 0.2662 \\
		& 288.65983 & -59.463317 & 0.2627 \\
		& 288.65554 & -59.472132 & 0.2635 \\
		& 288.67541 & -59.491637 & 0.2680 \\
RXCJ2006.4-1912 & & & \\
                & 301.57594 & -19.19882  & 0.2798 \\
		& 301.59097 & -19.207681 & 0.2772 \\
		& 301.60487 & -19.204003 & 0.2694 \\
		& 301.61177 & -19.205136 & 0.2769 \\
		& 301.63241 & -19.210241 & 0.2759 \\
RXCJ2041.2-1324 & & & \\
                & 310.30095 & -13.406979 & 0.3328 \\
		& 310.30239 & -13.407497 & 0.3311 \\
		& 310.2907  & -13.389369 & 0.3302 \\
		& 310.32768 & -13.423979 & 0.3295 \\
		& 310.31764 & -13.412583 & 0.3172 \\
RXCJ2049.9-3029 & & & \\
                & 312.45369 & -30.504551 & 0.2088 \\
		& 312.44101 & -30.498744 & 0.1998 \\
		& 312.45211 & -30.495267 & 0.2064 \\
		& 312.45157 & -30.474272 & 0.2143 \\
RXCJ2126.1-3202 & & & \\
                & 321.53756 & -32.038193 & 0.2805 & E(0.2807)\\
		& 321.54668 & -32.032845 & 0.2788 \\
RXCJ2149.0-3228 & & & \\
                & 327.21805 & -32.472663 & 0.2285 \\
		& 327.23078 & -32.477076 & 0.2304 \\
		& 327.25531 & -32.485818 & 0.2326 \\
		& 327.27583 & -32.476373 & 0.2333 \\
RXCJ2149.9-1859 & & & \\
                & 327.36016 & -19.031057 & 0.2852 \\
                & 327.36666 & -19.021321 & 0.2820 \\
RXCJ2308.3-0155 & & & \\
                & 347.05782 & -1.9312007 & 0.3101 \\
		& 347.06292 & -1.9292463 & 0.3046 \\
		& 347.06925 & -1.9297486 & 0.3030 \\
		& 347.10146 & -1.9380196 & 0.2971 \\
RXCJ2317.9-3534 & & & \\
                & 349.52696 & -35.624088 & 0.2041 \\
		& 349.49957 & -35.608131 & 0.1949 \\
		& 349.53337 & -35.60149 & 0.2048 \\
		& 349.49667 & -35.585751 & 0.1939 \\
                 & 349.65900 & -34.3772 & 0.1965 & 2dFGRS S600Z062(C) \\
RXCJ2351.0-1954 & & & \\
                & 357.76166 & -19.948692 & 0.2483 \\
		& 357.77012 & -19.940193 & 0.2471 \\
\end{longtable}
\end{document}